\documentclass[lettersize,journal]{IEEEtran}
\usepackage{amsmath,amsfonts}
\usepackage{algorithmic}
\usepackage{algorithm}
\usepackage{array}
\usepackage[caption=false,font=normalsize,labelfont=sf,textfont=sf]{subfig}
\usepackage{textcomp}
\usepackage{stfloats}
\usepackage{url}
\usepackage{hyperref}
\usepackage{verbatim}
\usepackage{graphicx}
\usepackage{cite}
\usepackage{multirow,enumitem}
\usepackage[mathscr]{eucal}
\usepackage{booktabs}
\usepackage{footmisc}
\usepackage{xcolor}
\usepackage{bbding}
\newcommand{\yy}{\boldsymbol{y}}

\newcommand{\system}{CARE}

\begin{document}

\title{Leveraging Content and Acoustic Representations for Speech Emotion Recognition}

\author{Soumya Dutta,~\IEEEmembership{Student Member,~IEEE,} and 
        Sriram Ganapathy,~\IEEEmembership{Senior Member,~IEEE}
\thanks{S. Dutta and S. Ganapathy are with Learning and Extraction and Acoustic Patterns (LEAP) laboratory, Electrical Engineering, Indian Institute of Science, Bangalore, India. This work was performed with grants received from Prime Ministers Research Fellowship (PMRF), Ministry of Education, India, Ministry of Information Technology (MEiTY) NLTM, India, Qualcomm Innovation Fellowship (QIF) as well as grants from British Telecom. }
\thanks{Manuscript received November xx, 2024.}}

\markboth{Journal of \LaTeX\ Class Files,~Vol.~14, No.~8, November~2024}%
{Shell \MakeLowercase{\textit{et al.}}: A Sample Article Using IEEEtran.cls for IEEE Journals}


\maketitle

\begin{abstract}
Speech emotion recognition (SER), the task of identifying the expression of emotion from spoken content, is challenging due to the difficulty in extracting  representations that capture emotional attributes. The scarcity of  labeled datasets further complicates the challenge where large models are prone to over-fitting.  
In this paper, we propose \system{} (\underline{C}ontent and \underline{A}coustic \underline{R}epresentations of \underline{E}motions), where we design a dual encoding scheme which emphasizes semantic and acoustic factors of speech. 
While the semantic encoder is trained using distillation from utterance-level text representations, the acoustic encoder is trained to predict low-level frame-wise features of the speech signal. 
The proposed dual encoding scheme is a base-sized model trained only on unsupervised raw  speech. With a simple light-weight classification model trained on the downstream task, we show that the \system{} embeddings  provide effective emotion recognition on  a variety of datasets. 
We  compare the proposal with several other self-supervised models as well as recent large-language model based approaches. 
In these evaluations, the proposed \system{}  is shown to be the best performing model based on average performance across   $8$ diverse datasets. 
We also conduct several ablation studies to analyze the importance of various design choices.
\end{abstract}


\section{Introduction}\label{sec:intro}
\IEEEPARstart{S}{peech} Emotion Recognition (SER) focuses on detecting the speaker's emotional state from the audio signal. Recognizing emotions in speech has significant applications across diverse fields, including human-computer interaction~\cite{pantic2005affective}, social media analysis~\cite{gaind2019emotion}, customer service call centers~\cite{li2019acoustic}, and mental health monitoring systems~\cite{ghosh2019emokey}. However, despite considerable progress, SER continues to pose challenges due to the complexity of human emotions and the inherent difficulties in effectively capturing them from limited labeled datasets.

Traditionally, SER systems have relied on various acoustic properties of speech signals. Lieberman et al.\cite{lieberman1962some}  emphasize the role of pitch contour in emotion analysis, while additional acoustic features, including energy, intensity, and speaking rate, were  recognized as  indicators of emotional class \cite{petrushin1999emotion}. The features identified through the Interspeech para-linguistic challenges were rich in emotional properties while being  high-dimensional~\cite{schuller2009interspeech, schuller2013interspeech}.  Eyben et al.~\cite{eyben2015geneva} introduced a minimalist feature set to address this dimensionality issue. 

In recent years, the network architectures in SER commonly include convolutional neural networks (CNN)\cite{yenigalla2018speech, dutta2022multimodal}, long short-term memory (LSTM) networks\cite{hsiao2018effective}, and transformer models~\cite{kumar2022speech}. While these models perform well on the specific datasets, they often struggle to generalize across diverse datasets. In such settings,  self-supervised learning (SSL) models have emerged as a promising solution.
Notable examples of SSL approaches include wav2vec 2.0~\cite{baevski2020wav2vec}, HuBERT~\cite{hsu2021hubert}, and WavLM~\cite{chen2022wavlm}. These models are engineered to capture speech patterns similar to textual models like BERT~\cite{devlin2019bert}. Although trained on neutral speech data, these models have demonstrated encouraging results in emotion recognition tasks~\cite{chen2023exploring,dutta2023hcam}. 
The emotion recognition performance may be further enhanced by training these models with emotion-aware self-supervised objectives. 
Two  recent examples are Vesper~\cite{chen2024vesper} and emotion2vec~\cite{ma2023emotion2vec}. 
However, emotion in speech is also shaped by its semantic content~\cite{li2023asr}. For instance, identifying emotions from text transcripts is often   more effective than interpreting them   from raw audio~\cite{dutta2023hcam}. The integration of speech content during the pre-training phase of SER models remains an under-explored yet promising area of research.

In this work, we introduce a self-supervised model for speech emotion recognition (SER) called \textbf{C}ontent and \textbf{A}coustic \textbf{R}epresentations of \textbf{E}motions (CARE). To the best of our knowledge, our approach  is the first effort to pre-train a self-supervised  model that integrates both semantic and acoustic components of speech. CARE leverages a dual encoding framework for processing speech signals: a semantic encoder, which aligns speech representations with sentence-level transcripts, and a non-semantic encoder, which aligns speech representations with low-level acoustic features from the PASE+ model~\cite{ravanelli2020multi}. The outputs of both encoders are combined, and a lightweight classification head is then trained to perform emotion recognition.
The key contributions are :-

    

\begin{itemize}
    \item Proposing a novel self-supervised model for speech emotion recognition (SER) consisting of dual encoders: a semantic encoder and an acoustic encoder.  
    \item Developing an   adaptation strategy for aligning pre-trained text models with speech inputs by convolutional  adapters. 
    \item Experimenting on $8$ benchmark speech datasets with diverse  tasks, showcasing the effectiveness of \system{}.
    \item Identifying the  individual  and collective impact of semantic and acoustic representations for emotion recognition. 
\end{itemize}

\section{Related work}\label{sec:related}

\subsection{Audio Feature Extraction for SER}
Recently, deep learning-based representations have gained popularity as low-level acoustic features. Notable examples include the SincNet architecture by Ravanelli et al.\cite{ravanelli2018speaker} and interpretable Gaussian filters by Agrawal et al.\cite{agrawal2020interpretable}. The LEAF front-end~\cite{zeghidour2021leaf}, was utilized by Dutta et al.~\cite{dutta2022multimodal} for speech emotion classification. Typically, these models require end-to-end training of both feature extractors and classifiers. In contrast, the proposed \system{} architecture is a self-supervised model designed to generalize across diverse datasets.
\subsection{Self-supervision for SER}
One of the earliest self-supervised model for the task of speech emotion recognition was proposed by Pascual et al.~\cite{pascual2019learning}. This consisted of processing a speech signal by the SincNet model~\cite{ravanelli2018speaker} followed by trainable convolutional blocks to predict a number of speech features such as the waveform, mel-frequency cepstral coefficients (MFCCs), pitch etc. Ravanelli et al.~\cite{ravanelli2020multi} further modified this model by adding more self-supervised tasks such as predicting FBANK and Gammatone features~\cite{schluter2007gammatone} to develop the PASE+ model. 

Among the general purpose speech SSL models that were proposed over the years, WavLM~\cite{chen2022wavlm}, was shown to outperform other models such as HuBERT~\cite{hsu2021hubert} and wav2vec2.0~\cite{baevski2020wav2vec} for emotion recognition.  Vesper~\cite{chen2024vesper} used a modified masking strategy to emphasize high pitch/energy regions of speech—known indicators of emotion—and derived targets for these masked regions from a WavLM teacher model.   A similar strategy was employed by Ma et al. in emotion2vec~\cite{ma2023emotion2vec}, which  utilized a pre-trained data2vec model as the teacher. Emotion2vec~\cite{ma2023emotion2vec} also learns a global embedding to enhance SER performance. In contrast, the proposed \system{} model integrates semantic content along with acoustic features. 

\subsection{\textcolor{black}{Multimodal Emotion Recognition}}
\textcolor{black}{The use of speech signals alongside text transcripts for multimodal emotion recognition has been explored in several prior works~\cite{dutta2022multimodal, fan2024leveraging, hu2024recent}. These approaches typically involve separate modeling of the two modalities, followed by a fusion stage.    In contrast, \system{} is designed to  model  the semantic and acoustic properties of speech with the uni-modal input.
}

\subsection{Speech-text Aligned Representations}
\begin{figure}
    \centering
    \includegraphics[width=0.6\textwidth,trim={3cm 4cm 7cm 3.5cm},clip]{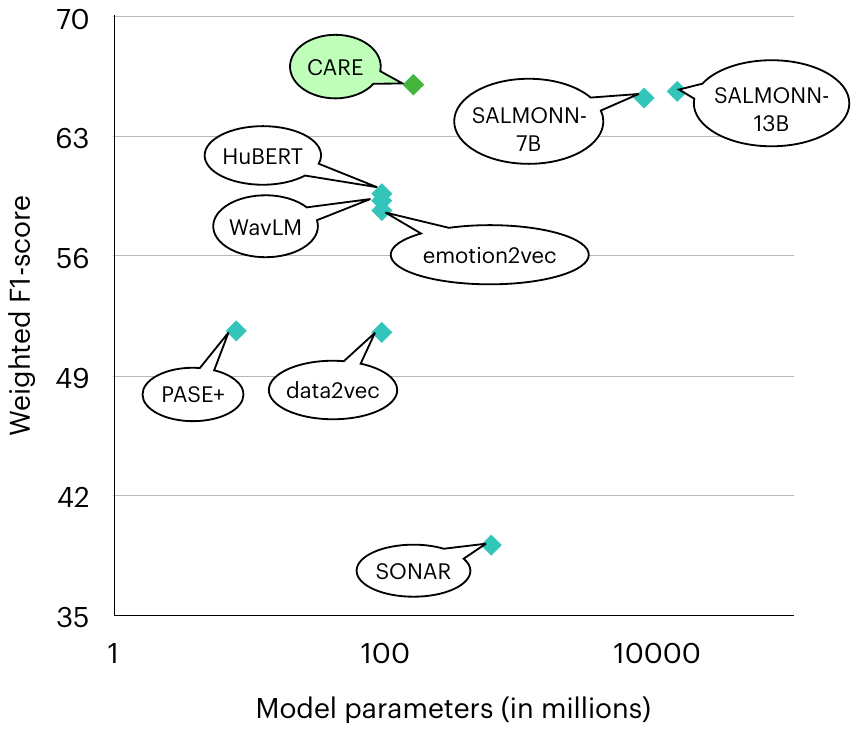}
    \vspace{-0.45in}
    \caption{Scatter plot of  inference model size (parameters in millions) versus the SER performance (average weighted F1-score over $8$ datasets). \system{} is seen to achieve a better trade-off compared to the existing solutions. For more details, refer Tables~\ref{tab:results-cat-oth} and ~\ref{tab:results-cat} and the associated discussions.}.
    \label{fig:compare}
    \vspace{-0.1in}
\end{figure}

The alignment of speech and text modalities has received renewed attention for speech representation learning. The SONAR model \cite{duquenne2023sonar} aligns a speech encoder with  textual representations at the utterance level. 
With the increasing prominence of large language models (LLMs), recent approaches have integrated speech encoders with LLMs.   Notably, the SALMONN model by Tang et al.\cite{tang2023salmonn} introduced an audio encoder consisting of Whisper model and a  music encoder along with the LLaMA language model~\cite{touvron2023llama}.   Hu et al.\cite{hu2024wavllm} proposed WavLLM, combining Whisper and WavLM encoders with the LLaMA model. These LLM-based approaches harness aligned speech-text representations, enabling prompt-based applications. However, their substantial model sizes (e.g., $7$B parameters for SALMONN) present significant computational demands for both training and inference.
In contrast, \system{} achieves superior performance on various downstream datasets with a much smaller size of $160$M parameters.\\
\textbf{Summary}: The landscape of various SER methods is summarized in Fig.~\ref{fig:compare}. We highlight a clear gap in current modeling frameworks: models either prioritize efficiency with limited performance (those in the lower end of the x-axis), or focus on maximizing performance with increased memory and compute requirements (typically based on LLMs). To address this gap, we propose \system{}, that combines the computational efficiency of smaller models with the high performance of large-scale systems, thereby providing a superior trade-off between efficiency and performance.

\begin{figure*}
    \centering
    \includegraphics[width=0.8\textwidth,trim={1.5cm 5cm 0cm 2.5cm},clip]{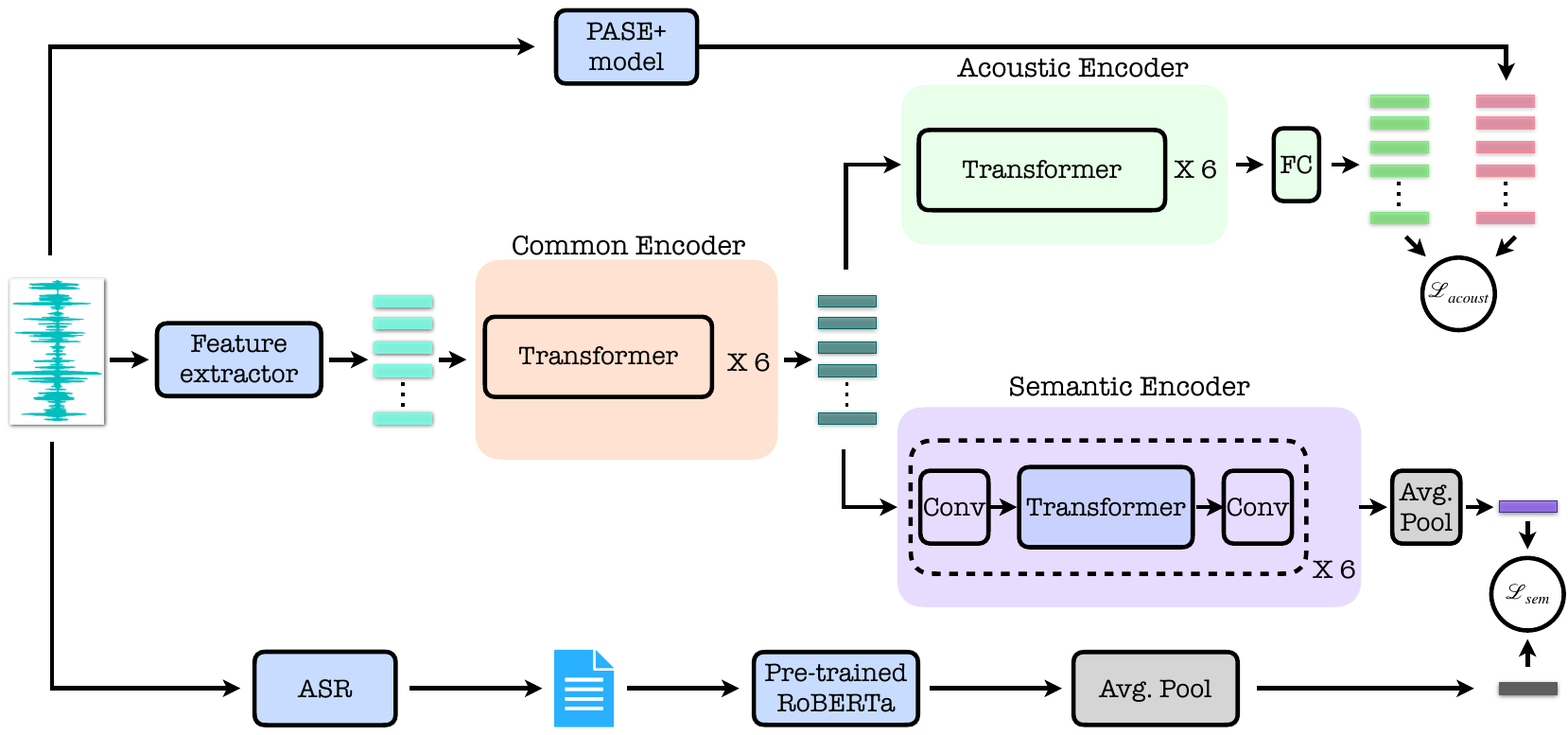}
    \caption{Block diagram of the proposed \system{} model.   The acoustic encoder of the model is trained with  PASE+ features as targets. Blocks in blue indicate either frozen components or those with no learnable parameters. For the semantic encoder the transformer layers are frozen while the convolutional adapters are trained. As the dimension of the output from the acoustic encoder is $768$, a FC layer is attached to match the PASE+ feature dimension of $256$. This FC layer and the average pool block after the semantic encoder are not used during inference. }
    \label{fig:entire model}
    \vspace{-0.1in}
\end{figure*}
\section{Proposed Approach}\label{sec:method}

\subsection{Background}

\subsubsection{RoBERTa}
One of the significant contributions in creating a text representation model was proposed by Devlin et al.~\cite{devlin2019bert}.
Liu et. al \cite{liu2019roberta} trained this architecture on a larger corpus of textual data without the next sentence prediction task. This pre-trained model, known as robust optimized BERT approach (RoBERTa), was shown to outperform BERT in a number of downstream tasks.

\subsection{\system{} Model}
 
We propose a dual encoding scheme (semantic and acoustic encoders) to  process the speech signal through distinct supervisory signals suited to their respective objectives. The chosen supervision for each encoder is detailed as follows:\\
\textbf{Semantic supervision:}
We do not assume the availability of   ground-truth text transcripts for the pre-training data.  In such a scenario, pre-trained automatic speech recognition (ASR) systems (Whisper-large-v3~\cite{radford2023robust}) offer an alternative for generating these transcripts. 
Typically, ASR systems have been shown  to exhibit higher word error rates (WER) on emotional speech compared to neutral speech datasets~\cite{li2023asr}. 
Podcast recordings, on the other hand,  provide sufficiently long context and offer a broad content variety suitable for pre-training the semantic encoder. Specifically, we observe a WER of $12.53\%$, which may be reasonable for SER tasks.

Since the semantic encoder’s purpose is to align the speech signal with its content to facilitate emotion recognition, an ASR-style alignment loss could be applied. However, a sentence-level representation for text is more appropriate for the task of emotion recognition as established by Fan et al.~\cite{fan2022sentiment}. Therefore, we extract contextual word-level embeddings from the transcripts using a pre-trained RoBERTa model~\cite{liu2019roberta} and mean-pool these embeddings to obtain a single feature vector representing the entire transcript. These utterance-level embeddings serve as the supervisory signal, or ``teacher'', for the semantic encoder in our \system{} model. We denote these utterance-level embeddings by $\yy_{text}$.\\
\noindent \textbf{Acoustic Supervision:}
In prior works, mean-pooled   representations have shown to encode characteristics like speaker identity, accent, and language~\cite{krishna2024towards}. However,  we speculate that 
emotion in speech is often contained in fine-grained acoustic attributes  such as pitch, rhythm, and their modulations~\cite{petrushin1999emotion}. Thus, a frame-level target is chosen for the acoustic encoder.

A direct approach for the frame level acoustic targets would involve masking parts of the speech signal and reconstructing them.
However, prior works show that random masking is less effective for emotion recognition than selectively masking high-energy or high-pitch regions, as demonstrated by Chen et al.~\cite{chen2024vesper}. 
Based on these observations, we choose to predict PASE+ features, which encompass filter-bank energies, pitch, and other low-level descriptors essential for capturing emotion. Specifically, we use frame-level PASE+ features with $256$ dimensions as targets for the acoustic encoder in our \system{} model. These features are down-sampled by a factor of $2$, producing target descriptors at a frequency of $50$ Hz. We denote the acoustic targets from the PASE+ model by $\yy_{pase}$.

\subsubsection{\textbf{Model Architecture}}
The speech signal is first processed through a series of convolutional layers designed to produce frame representations every $20$ ms. These are followed by a stack of six transformer layers, forming the common encoder that serves both the acoustic and semantic encoder pathways in the proposed model.


The semantic encoder is designed to align the speech representations with its corresponding generated transcript. This encoder consists of six transformer layers which are initialized with the weights from a pre-trained text representation model. Being trained with textual data, the transformer layers in the semantic encoder do not generalize to speech representations. To address this, we propose a novel adaptation strategy by introducing two 1D-convolutional blocks—one placed before and one after each transformer layer. 

The first block adjusts the speech representations from the common encoder to align them with the internal representations expected by the text-based model. The second block refines these representations post-transformer processing. Additionally, following established practice for processing speech in text models~\cite{yu2024connecting, fathullah2024prompting, tang2023salmonn}, the time resolution of the speech sequence is reduced before processing by the transformer layers in the semantic encoder. Specifically, the convolutional block preceding each transformer layer down-samples the sequence length by a factor of three, while the block following it up-samples it by the same factor. Each convolutional block consists of a single convolutional layer, with a kernel size of $5$, and input and output channels set to $768$ in order to match the dimension of the pre-trained transformer layers. While adaptation of speech SSL models with convolution layers has been explored in prior works~\cite{li2023evaluating, kim2024convolution}, adapting pre-trained text models for speech tasks, using convolutional adapters, is explored for the first time in this work. Importantly, the transformer layers themselves are not updated during training. Finally, the semantic encoder’s output representations are average-pooled to produce an utterance-level representation. \par

The acoustic encoder also consists of six transformer layers, with its output subsequently mapped to $256$ dimensions, using a fully-connected layer, to match the PASE+ feature targets. Figure \ref{fig:entire model} provides a block diagram of the \system{} model.

 \subsubsection{\textbf{Loss}}  A semantic loss, $L_{sem}$ and a frame-level acoustic loss, $L_{acoust}$, are employed for training the semantic and acoustic encoders, respectively. 
 Denoting the semantic supervision by $\yy_{text}$ and the output from the semantic encoder as $\hat{\yy}_{sem}$, the semantic loss is the mean square error (MSE) loss:
\begin{equation}\label{eq:utt}
    L_{sem.} = \frac{1}{N}\sum_{i=1}^{N}||\yy_{text}^{i} - \hat{\yy}_{sem}^{i}||_{2}^{2}
\end{equation}
where $N$ denotes the batch size.\par
For the frame level loss, let $\hat{\yy}_{acoust} \in \mathbb{R}^{N\times T \times D}$ denote the output of the acoustic encoder, where $N$, $T$ and $D$ denote the batch size, number of frames per utterance and the dimension of the representation,  respectively. The loss is defined as:
\begin{equation}\label{eq:frame}
    L_{acoust.} =  \frac{1}{NT}\sum_{i=1}^{N}\sum_{j=1}^{T}||\yy_{pase}^{ij} - \hat{\yy}_{acoust}^{ij}||_{2}^{2}
\end{equation}
where $\yy_{pase}$ denotes the acoustic target.
\textcolor{black}{The total loss during pre-training is given by}
\begin{equation}\label{eq:total}
    \textcolor{black}{L_{tot.} = L_{sem.} + \lambda L_{acoust.}}
\end{equation}
\textcolor{black}{where $\lambda$ is decided based on the validation performance.}
\subsubsection{\textbf{Inference}}
For evaluating the model across various downstream tasks, we adopt the paradigm proposed in the SUPERB benchmark~\cite{yang2021superb}. 
The outputs from each transformer layer of the acoustic encoder are concatenated with the outputs from the convolution block following each transformer layer in the semantic encoder. These are then combined with layer-wise outputs from the common encoder and the convolutional feature extractor. This process yields a total of $13$ layer representations—one from the convolutional feature extractor, six from the common encoder, and six from the concatenated semantic and acoustic encoders. A convex combination of these layer representations is then fed into a classification head. It is to be noted that, during inference, the fully-connected layer in the acoustic encoder and the average pooling block in the semantic encoder are not used.

In this setup, the only learnable parameters for the downstream tasks are the weights for the convex combination and those of the lightweight classification head.

\section{Experiments and Results}
\subsection{Pre-training}
 The MSP-PODCAST corpus~\cite{lotfian2017building} is used for the task of pre-training. A total of $149,307$ samples amounting to $230$ hours of emotional speech data are used. Out of these, $80\%$ of the data is randomly chosen as the training set while the remaining $20\%$ serves as the validation set. The Whisper-large-v3 model is used for generating the transcripts (the WER observed is $12.53\%$), while the pre-trained RoBERTa model is used for encoding the transcripts. The common encoder is initialized with first $6$ layers of the WavLM-base model, while the acoustic encoder is initialized with the last $6$ layers of the same. The convolutional feature extractor is also initialized from the WavLM-base model. The $6$ transformer layers of the semantic encoder are initialized with the weights of the last $6$ layers of a pre-trained RoBERTa base model, while the convolutional adapters are randomly initialized.
\begin{table}[t!]
\centering
\caption{Summary of the evaluation datasets. Class balanced denotes if the datasets are balanced across classes. The last column indicates if the training/test data have common speakers. For all the datasets, we perform $5$-fold evaluation.}\label{tab:datasets}
    \resizebox{\columnwidth}{!}{%

\begin{tabular}{l|c|c|c|c|c|c}
\toprule
Datasets  & \begin{tabular}[c]{@{}c@{}}\# Train\\ Utt. \end{tabular} & \begin{tabular}[c]{@{}c@{}}\# Val.\\ Utt. \end{tabular} & \begin{tabular}[c]{@{}c@{}}\# Test\\ Utt. \end{tabular} & \# Classes & \begin{tabular}[c]{@{}c@{}} Class\\ Bal. \end{tabular}&\begin{tabular}[c]{@{}c@{}}Spkr.\\ Ind. \end{tabular} \\ \midrule
  
IEMOCAP-4 \cite{busso2008iemocap} & $4425$ & $1102$ & $1102$ & 4 & \XSolid & \Checkmark \\ \midrule
IEMOCAP-6 \cite{busso2008iemocap} & $5947$ & $1487$ & $1387$ & $6$ & \XSolid & \Checkmark \\ \midrule
MELD \cite{poria2019meld} & $9988$ & $1108$ & $2610$ & $7$ & \XSolid & \XSolid \\ \midrule
CMU-MOSI \cite{zadeh2016mosi} & $1188$ & $325$ & $686$ & $2$ & \Checkmark & \Checkmark \\ \midrule
DAIC-WOZ \cite{gratch2014distress} & $6003$ & $2097$ & $2097$ & $2$ & \Checkmark& \Checkmark \\ \midrule
RAVDESS-Song \cite{livingstone2018ryerson} & $704$ & $132$ &$176$  &$6$  &\Checkmark& \Checkmark  \\ \midrule
CaFE \cite{gournay2018canadian}  & $624$ & $156$ &$156$  & $7$ & \Checkmark& \Checkmark  \\ \midrule
EmoDB \cite{burkhardt2005database} & $324$ & $105$ &$106$  &$7$  &\Checkmark& \Checkmark  \\  
\bottomrule
\end{tabular}}
\vspace{-0.1in}
\end{table}
\vspace{-0.1in}

\subsection{Downstream Tasks}
A summary of the different datasets used for evaluation is mentioned in Table~\ref{tab:datasets}.
\subsubsection{IEMOCAP}
The IEMOCAP dataset consists of $151$ video recordings split into 5 sessions. Each of these sessions is a conversation between a pair of subjects. Each recording is split into multiple utterances. There are a total of $10,039$  utterances, each of which is labeled by human annotators as belonging to one of the $10$ emotions -  ``angry'', ``happy'', ``sad'', ``neutral'', ``frustrated'', ``excited'', ``fearful'', ``surprised'', ``disgusted'' or ``other''. Keeping in line with previous works, we do a four-way classification task where we consider ``angry'', ``happy'', ``sad'', ``neutral'' and ``excited'' categories (with ``excited'' and ``happy'' categories merged). We also have a separate setting of $6$ emotional classes~\cite{majumder2019dialoguernn}. The first $6$ of the $10$ emotion classes are considered for this setting.
\subsubsection{MELD}
The MELD dataset~\cite{poria2019meld} is a dataset created from video clippings of the popular TV show, ``Friends''.  A seven way classification task is performed on this dataset, with each utterance being labeled as one of the $7$ emotions - ``angry'', ``sad'', ``joy'', ``neutral'', ``fear'', ``surprise'' or ``disgust''.
\subsubsection{CMU-MOSI}
The CMU-MOSI dataset~\cite{zadeh2016mosi} has a total of $2199$ utterances. Each utterance is labeled in the range of $[-3, 3]$. Following previous works, we treat this as a binary classification problem with utterances having sentiment values in the range $[-3, 0)$ being classified as negative sentiment and those with values in the range $[0, 3]$ considered as positive sentiment. The dataset partitioning  follows a prior work~\cite{poria2017context}.
%
\subsubsection{DAIC-WOZ}
The DAIC-WOZ dataset~\cite{gratch2014distress} is a benchmark dataset for depression detection, consisting of $189$ clinical interviews between the patient and the interviewer. Out of these $189$ interviews, $107$ are part of the training set while $35$ interviews are part of the development subset. The dataset suffers from a data imbalance problem, with only $30$   interviews labeled as ``depressed'' in the train  set . In order to increase the balance, we follow~\cite{wu2023self} and extract $100$ utterances randomly from each interview, which is labeled as depressed, while only $39$ utterances are selected for interviews classified as ``normal''. 
The utterances from each interview are chosen randomly for the $5$ splits. Following prior work~\cite{shen2022automatic,ravi2022step,wu2022climate,wu2023self}, we report results on the development set of this dataset.
\subsubsection{RAVDESS-Song}
The RAVDESS-Song dataset~\cite{livingstone2018ryerson} has a total of $1012$ song recordings by $23$ different singers. Each recording in this dataset is sung in one of six different emotions, namely, ``neutral'', ``calm'', ``happy'', ``sad'', ``angry'' and ``fear''. We conduct a speaker independent evaluation for this dataset, and create $5$ different splits. For each split, we keep recordings from $16$ singers for training, while recordings from  $3$ separate singers are used for validation. The recordings from the remaining $4$ speakers are used for evaluation. 
\subsubsection{CaFE}
The CaFE dataset~\cite{gournay2018canadian} is a Canadian French emotional dataset consisting of $936$ utterances spoken by $12$   speakers. Each utterance in this dataset is categorized as one of the seven emotions - ``neutral'', ``angry'', ``disgust'', ``sad'', ``surprise'', ``fear'' and ``happy''. Similar to RAVDESS-Song, we create $5$ speaker independent splits for this dataset. The utterances belonging to $8$   speakers are used for the training,  while the remaining $4$ speakers are used for validation and testing equally. The speakers used for train, validation and test are chosen randomly for the $5$ splits.
\subsubsection{EmoDB}
The EmoDB dataset~\cite{burkhardt2005database} has a total of $535$ utterances spoken by 10 different speakers for the task of emotion recognition in German. Each utterance in this dataset is categorized as one of the seven emotions - ``neutral'', ``angry'', ``disgust'', ``sad'', ``boredom'', ``fear'' and ``joy''. Similar to RAVDESS-Song and CaFE, we create $5$ different speaker independent splits for this dataset. The utterances belonging to $6$ different speakers are chosen for  training  while the remaining $4$ speakers are used for validation and testing equally. The speakers used for train, validation and testing are chosen randomly for each of the $5$ splits.
\subsection{Loss and Evaluation Metrics for Downstream Tasks}
The cross-entropy loss is used for training the downstream model weights (the convex combination weights and the lightweight classification head parameters). 
For testing, we use the weighted F1-score as the evaluation metric as many of the datasets are class-imbalanced  (Table~\ref{tab:datasets}). Denoting the F1 score of class $c$ with $N_c$ samples, by $F1_c$, the weighted F1-score is
\begin{equation}
    WF1 = \frac{1}{\sum_{c=1}^{C}N_c}\sum_{c=1}^{C}N_c\times F1_c
\end{equation}
\textcolor{black}{We also report the unweighted average recall (UAR) for all cases, which is the mean of the class-wise recall scores.}
\subsection{Implementation Details}
\subsubsection{\textbf{Pre-training}} During pre-training, all the speech utterances from MSP-PODCAST are padded or randomly cropped to a duration of $5$ seconds. The model is trained with a learning rate of $1e$-$5$ and a batch size of $128$ with AdamW~\cite{loshchilov2017decoupled} as the optimizer. The model is trained for a total of $200,000$ steps and the best model parameters based on validation set performance are  chosen for evaluation of downstream datasets. \textcolor{black}{We experiment with different values of $\lambda$ (Eq.~\ref{eq:total}) to balance the two losses during pre-training. Setting $\lambda=0.1$ results in degraded performance, while increasing it to $\lambda=10$ does not yield any significant improvement over $\lambda=1$. Therefore, we fix $\lambda=1$  for all the subsequent  experiments.}\par
\subsubsection{\textbf{Fine-tuning and evaluation}} For the downstream task training,  the speech signals are cropped to a maximum duration of $30$ seconds or padded to a minimum duration of $1$ second. For the depression detection dataset, DAIC-WOZ, each speech segment has a duration of $10$ seconds~\cite{wu2023self}.

Each layer output in the common, semantic, and acoustic encoders has a dimensionality of $T\times768$, where $T$ denotes the number of frames in the speech signal, sampled at $50$Hz. For the \system{} model, as outputs from the $6$ semantic and acoustic encoder layers are concatenated, the combined output dimension is $6\times T \times1536$. To align with this dimensionality, the output from the convolutional feature extractor and the common encoder's $6$ layers are duplicated to yield features of dimension $7 \times T \times 1536$. Representations from these $13$ layers are combined through a convex combination approach with learnable weights producing features of dimension $T \times 1536$. Following this, features are mean-pooled along the temporal dimension, producing a single $1536$-dimensional vector per audio file. This is input into a classification head consisting of a two-layer feed-forward neural network that employs ReLU activation~\cite{nair2010rectified}. Only the weights for the convex combination of layer representations and those in the two-layer fully connected classification head are trained on each downstream dataset, consistent with the SUPERB framework~\cite{yang2021superb}.

We use a batch size of $32$ with a learning rate of $1e$-$4$ and train the model for   $50$ epochs. The hidden dimension of the two-layer classification head is set to be $256$. The AdamW optimizer is used here as well. All the models, including the \system{} and the baseline systems, utilize the same classification backend. Thus, the  design allows fair comparison of the different representations.
\footnote{Code available at \url{https://github.com/iiscleap/CARE}.}\par


\begin{table*}[t!]
\centering
\caption{Comparison with other works for downstream datasets. $^{\#}$  models which include downstream dataset in pre-training. Results in \textbf{bold}, \underline{underlined} indicate the best  and the second-best model, respectively. All numbers are weighted F1-scores computed over $5$ random initializations (mean and standard deviation shown). \textcolor{black}{The Unweighted Average Recall is also shown in brackets.}\label{tab:results-cat-oth}  }
    \resizebox{0.99\textwidth}{!}{%
\begin{tabular}{@{}l|c|c|c|c|c|c|c||c@{}}
\toprule
\multicolumn{1}{l|}{\multirow{2}{*}{Datasets}} &WavLM \cite{chen2022wavlm} &HuBERT \cite{hsu2021hubert} & data2vec \cite{baevski2022data2vec} & emotion2vec \cite{ma2023emotion2vec} & SONAR \cite{duquenne2023sonar} & \multicolumn{2}{c||}{SALMONN \cite{tang2023salmonn}}  & \system{} \\ 
& \textbf{Params}:$94$M & \textbf{Params}:$94$M & \textbf{Params}:$94$M 
& \textbf{Params}:$94$M & \textbf{Params}:$600$M & \textbf{Params}:$7$B & \textbf{Params}:$13$B & \textbf{Params}:$160$M \\ \midrule
IEMOCAP-4 & $65.9^{\pm0.5}\textcolor{black}{(67.2)}$ & $65.0^{\pm0.2}\textcolor{black}{(68.0)}$ & $62.7^{\pm0.7}\textcolor{black}{(64.0)}$ & $67.5^{\pm0.6\#}\textcolor{black}{(69.0)}$ & $59.4^{\pm0.4}\textcolor{black}{(61.0)}$ & $\mathbf{75.8}^{\pm0.6\#}\textcolor{black}{(76.9)}$ & $\underline{72.9}^{\pm2.3\#}\textcolor{black}{(74.9)}$ & $69.4^{\pm0.5}\textcolor{black}{(70.1)}$ \\ \midrule
IEMOCAP-6 & $51.7^{\pm0.5}\textcolor{black}{(48.4)}$ & $50.7^{\pm0.9}\textcolor{black}{(46.5)}$ & $46.0^{\pm0.4}\textcolor{black}{(42.3)}$ & $54.1^{\pm0.6\#}\textcolor{black}{(51.9)}$ & $43.5^{\pm0.2}\textcolor{black}{(41.0)}$ & $\mathbf{59.3}^{\pm1.4\#}\textcolor{black}{(55.7)}$ & $\underline{58.1}^{\pm1.6\#}\textcolor{black}{(55.2)}$ & $55.0^{\pm0.4}\textcolor{black}{(52.1)}$ \\ \midrule
MELD &  $45.6^{\pm0.4}\textcolor{black}{(24.3)}$ & $45.3^{\pm0.6}\textcolor{black}{(24.0)}$ & $41.9^{\pm0.5}\textcolor{black}{(23.1)}$ & $47.6^{\pm0.3\#}\textcolor{black}{(27.4)}$ & $43.2^{\pm0.2}\textcolor{black}{(23.3)}$ & $\mathbf{53.3}^{\pm0.7}\textcolor{black}{(33.4)}$  & $\underline{52.6}^{\pm0.4}\textcolor{black}{(32.8)}$ & $48.1^{\pm0.8}\textcolor{black}{(28.8)}$ \\ \midrule
CMU-MOSI &  $64.1^{\pm0.8}\textcolor{black}{(64.2)}$ & $62.5^{\pm0.6}\textcolor{black}{(62.5)}$ & $59.7^{\pm0.4}\textcolor{black}{(58.9)}$ & $66.5^{\pm0.6}\textcolor{black}{(65.9)}$ & $\underline{74.6}^{\pm0.3}\textcolor{black}{(73.9)}$ & $\mathbf{78.0}^{\pm0.7}\textcolor{black}{(77.0)}$ & $72.8^{\pm1.0}\textcolor{black}{(72.0)}$ & $66.7^{\pm1.0}\textcolor{black}{(66.2)}$ \\ \midrule
DAIC-WOZ &  $63.2^{\pm1.5}\textcolor{black}{(61.5)}$ & $65.9^{\pm2.0}\textcolor{black}{(61.9)}$ & $\underline{67.8}^{\pm1.4}\textcolor{black}{(65.7)}$ & $61.6^{\pm0.7}\textcolor{black}{(61.0)}$ & $64.3^{\pm0.4}\textcolor{black}{(63.7)}$ & $62.6^{\pm3.4}\textcolor{black}{(60.4)}$ & $64.7^{\pm3.0}\textcolor{black}{(61.1)}$ & $\mathbf{68.5}^{\pm2.1}\textcolor{black}{(67.1)}$ \\ \midrule
RAVDESS &  $50.5^{\pm3.6}\textcolor{black}{(49.1)}$ & $\underline{53.5}^{\pm1.1}\textcolor{black}{(55.7)}$ & $38.5^{\pm5.2}\textcolor{black}{(40.8)}$ & $48.5^{\pm1.0}\textcolor{black}{(51.0)}$ & $11.8^{\pm2.0}\textcolor{black}{(10.8)}$ & $50.2^{\pm1.3}\textcolor{black}{(54.2)}$ & $51.9^{\pm3.6}\textcolor{black}{(53.4)}$ & $\mathbf{60.1}^{\pm1.6}\textcolor{black}{(62.0)}$ \\ \midrule
CaFE & $66.6^{\pm2.6}\textcolor{black}{(69.0)}$ &$66.5^{\pm4.5}\textcolor{black}{(69.1)}$ & $48.8^{\pm4.3}\textcolor{black}{(51.0)}$ & $59.3^{\pm3.8}\textcolor{black}{(62.6)}$ & $5.7^{\pm1.4}\textcolor{black}{(7.1)}$ & $59.9^{\pm2.0}\textcolor{black}{(62.8)}$ & $\underline{69.8}^{\pm3.3}\textcolor{black}{(71.4)}$ & $\mathbf{77.0}^{\pm1.5}\textcolor{black}{(78.1)}$ \\ \midrule
EmoDB & $66.5^{\pm4.8}\textcolor{black}{(68.5)}$ & $66.9^{\pm3.9}\textcolor{black}{(68.2)}$&  $48.9^{\pm3.1}\textcolor{black}{(49.9)}$ &  $64.4^{\pm2.6}\textcolor{black}{(66.7)}$ & $10.2^{\pm2.0}\textcolor{black}{(12.6)}$ & $\underline{82.8}^{\pm2.9}\textcolor{black}{(85.3)}$ & $82.2^{\pm4.0}\textcolor{black}{(84.1)}$ & $\mathbf{83.4}^{\pm2.0}\textcolor{black}{(83.9)}$ \\ \midrule \midrule
Avg. & $59.3\textcolor{black}{(56.5)}$ & $59.5\textcolor{black}{(57.0)}$ & $51.8\textcolor{black}{(49.5)}$ &  $58.7\textcolor{black}{(56.9)}$ & $39.1\textcolor{black}{(36.7)}$ & $65.2\textcolor{black}{(63.2)}$& $\underline{65.6}\textcolor{black}{(63.1)}$ & $\mathbf{66.0}\textcolor{black}{(63.5)}$ \\ \bottomrule
\end{tabular}}
\vspace{-0.1in}
\end{table*}

\subsection{Performance of \system{}}
The results on the $8$ downstream datasets using representations from the proposed \system{} model are shown in Table~\ref{tab:results-cat-oth}.  \textcolor{black}{These baseline models are categorized into two groups based on the number of parameters used during inference: base models (parameter size $<200$M), and large models ($>500$M), which also include LLM based models.} The following observations are made for each category:
\subsubsection{\textbf{Base models}}
We compare HuBERT \cite{hsu2021hubert}, WavLM \cite{hu2024wavllm}, data2vec \cite{baevski2022data2vec} and emotion2vec~\cite{ma2023emotion2vec} representations as the baseline models in this category. Among these baseline systems, the emotion2vec is also pre-trained on IEMOCAP and MELD datasets, partially explaining the improved results seen on the downstream tasks on these datasets.
While \system{} performs similar to emotion2vec on CMU-MOSI, it improves over all the base-sized models on other datasets. On the average, the proposed \system{} achieves a relative improvement of $15.6$\% over the best baseline model (HuBERT). 

\subsubsection{\textbf{Large models}}
SONAR~\cite{duquenne2023sonar} is selected as the speech encoder in this category. For the six English-based datasets, the pre-trained English speech encoder\footnote{\url{https://dl.fbaipublicfiles.com/SONAR/spenc.eng.pt}} is used, while the French and German speech encoders are utilized for the CaFE and EmoDB datasets, respectively. Similar to the \system{} backend, the layer representations from the SONAR encoder are linearly combined and the classification head is trained on the downstream task.
 Although SONAR has nearly four times the parameter size of \system{}, our proposed model outperforms SONAR across all datasets except CMU-MOSI. 

\subsubsection{\textbf{LLM based models}}
Two versions of SALMONN~\cite{tang2023salmonn} ($7$B and $13$B)\footnote{\url{https://huggingface.co/tsinghua-ee/SALMONN}} are considered as examples of LLM-based models. 
        These are typically applied in a zero-shot setting; however, due to variability in emotion classes across datasets, their zero-shot performance is inconsistent. E.g. while SALMONN-$13$B model achieves $68.75\%$ weighted F1-score on the IEMOCAP-4 dataset (on which it is trained), it achieves only $24.06\%$ for MELD. Thus, for fair comparison, the same framework used in \system{} and other baseline models is followed for the LLM based evaluations as well. The internal representations from all layers ($41$ layers for SALMONN $13$B and $33$ layers for SALMONN $7$B) are aggregated using a convex combination, and the classification head (similar to \system{}) is trained for each downstream dataset. 
    Similar to emotion2vec, SALMONN includes IEMOCAP in its pre-training, leading to superior performance on IEMOCAP-4 and IEMOCAP-6 compared to \system{}. The larger model size and extensive pre-training data allows SALMONN to outperform \system{} by $10\%$ and $34\%$ (relative improvements) on the MELD and CMU-MOSI datasets, respectively. 
      However, on the remaining four tasks, \system{} surpasses the SALMONN models, achieving relative improvements of $17\%$ and $24\%$ on the RAVDESS-song and CaFE datasets, respectively. Notably, though music datasets are used to pre-train SALMONN, \system{} emerges as the best model on the RAVDESS-Song dataset.\\
\noindent{\textbf{Key takeaways:}}
1) On average, \system{} emerges as the top-performing model across the eight datasets, surpassing even the SALMONN $13$B model, which has nearly $80$ times more parameters. 
Although LLM-based models show strengths in in-domain emotion recognition datasets, their performance declines on out-of-domain tasks, indicating limited generalizability across diverse tasks and multilingual emotional speech. 2) \system{}’s advantage over speech SSL models like WavLM, HuBERT, and data2vec is expected, given that these models are trained on non-emotional data (see  Sec.~\ref{sec:cont} for a related experiment). 3) Notably, \system{} outperforms the multilingual SONAR model on CaFE and EmoDB datasets although it is trained on English speech only. This showcases the generalizability of our pre-training technique to out-of-domain tasks in SER.
\subsection{Emotional Attribute Prediction}
The   emotion recognition can be posed as a regression problem, where valence, arousal and dominance of a particular utterance are predicted~\cite{parthasarathy2017jointly}. We use the MSP-IMPROV~\cite{busso2016msp} for this purpose. This is an audio-visual dataset that consists of $12$ actors eliciting a set of sentences in different emotions. The dataset consists of $8438$ utterances with valence, arousal and dominance values (ranging from $1$ to $5$). We split the dataset in $12$ parts, where  each part contains utterances corresponding to $10$ training speakers, while speech from the two other speakers are used for validating and testing the model. The performance is measured as the average over these $12$ parts.\par
We use the concordance correlation coefficient (CCC) as the metric. Denoting the mean, variance of ground truth by $\mu_g$, $\sigma_g^2$  and predicted scores by $\mu_p$, $\sigma_p^2$, the CCC is defined as 
\begin{equation}\label{eq:ccc}
    CCC = \frac{2\rho\sigma_g\sigma_p}{\sigma_p^{2} + \sigma_{g}^{2} + (\mu_g-\mu_p)^2}
\end{equation}
In Eq.~\ref{eq:ccc}, $\rho$ is the Pearson's correlation coefficient between the ground truth and the predicted scores. For training the downstream model, the representations from \system{} and other models are aggregated similar to the categorical datasets. This is followed by a two-layer regression head with $256$ as the hidden dimension and $3$ as the output dimension ($1$ for each of the three attributes). The objective is to increase the CCC between the ground truth and the predicted values for each of the dimensions of valence, arousal and dominance. 
The results for this dataset along with other baseline models are shown in Table.\ref{tab:results-dim-oth}. We note that for this task, the \system{} embeddings achieve the best results in terms of the valence and dominance attributes, while the performance  on arousal is marginally better for the SALMONN-$7$B model.
\begin{table}[t!]
\centering
\caption{Results for the MSP-IMPROV dataset. CCC stands for Concordance Correlation Coefficient while V, A, D stand for valence, arousal and dominance respectively.}\label{tab:results-dim-oth}
    \resizebox{0.7\columnwidth}{!}{%
\begin{tabular}{@{}l|l|l|l@{}}
\toprule
Method & CCC-V & CCC-A & CCC-D \\ \midrule
WavLM-base \cite{chen2022wavlm} & $0.51$ & $0.64$ & $0.47$ \\ \midrule
 emotion2vec \cite{ma2023emotion2vec} & $0.5$ & $0.61$ & $0.49$ \\ \midrule
 SALMONN-$7$B \cite{tang2023salmonn} & $0.53$ & $\mathbf{0.67}$ & $\underline{0.52}$ \\ \midrule
 SALMONN-$13$B \cite{tang2023salmonn} & $\underline{0.56}$ & $0.65$ & $0.51$ \\ \midrule
 \system{} & $\mathbf{0.57}$ & $\underline{0.66}$ & $\mathbf{0.53}$ \\ \bottomrule
\end{tabular}}
\end{table}
\section{Discussion}

\begin{table*}[t!]
\centering
\caption{Baseline results on the different downstream datasets in terms of weighted F1-score. All numbers are averaged over $5$ random initializations of the downstream network. We also show the results of the different components of \system{} in this table.  }\label{tab:results-cat}
    \resizebox{1.75\columnwidth}{!}{%
\begin{tabular}{@{}l|c|c|c|c||c|c|c@{}}
\toprule
\multicolumn{1}{l|}{\multirow{2}{*}{Datasets}} 
& PASE+ \cite{ravanelli2020multi} 
& \textcolor{black}{Whisper~\cite{radford2023robust}}
& \begin{tabular}[c]{@{}c@{}}Whisper \cite{radford2023robust}+\\ RoBERTa \cite{liu2019roberta}\end{tabular} 
& \begin{tabular}[c]{@{}c@{}}\textcolor{black}{Teacher-}\\\textcolor{black}{fusion}\end{tabular}
& \begin{tabular}[c]{@{}c@{}}Semantic +\\ Common Enc.\end{tabular}  
& \begin{tabular}[c]{@{}c@{}}Acoustic +\\ Common Enc.\end{tabular} 
& \system{} \\ 
 & \textbf{Params}:$8$M 
 & \textcolor{black}{\textbf{Params}:$800$M} 
 & \textbf{Params}:$1.6$B 
 & \textcolor{black}{\textbf{Params}:$1.6$B}
 & \textbf{Params}:$110$M 
 & \textbf{Params}:$94$M 
 & \textbf{Params}:$160$M \\ 
\midrule
IEMOCAP-4 \cite{busso2008iemocap} & $56.68$ & \textcolor{black}{$56.40$} & $61.97$ & \textcolor{black}{$69.49$} & $66.44$ & $65.91$ & $\mathbf{69.39}$ \\ 
\midrule
IEMOCAP-6 \cite{busso2008iemocap} & $41.38$ & \textcolor{black}{$40.62$} & $49.28$ & \textcolor{black}{$56.61$} & $53.05$ & $52.09$ & $\mathbf{55.02}$ \\ 
\midrule
MELD \cite{poria2019meld} & $35.86$ & \textcolor{black}{$40.11$} & $\mathbf{49.29}$ & \textcolor{black}{$49.72$} & $47.37$ & $46.98$ & $48.05$ \\ 
\midrule
CMU-MOSI \cite{zadeh2016mosi} & $50.69$ & \textcolor{black}{$55.60$} & $\mathbf{75.14}$ & \textcolor{black}{$74.12$} & $64.23$ & $64.17$ & $66.74$ \\ 
\midrule
DAIC-WOZ \cite{gratch2014distress, devault2014simsensei} & $66.84$ & \textcolor{black}{$62.08$} & $64.04$ & \textcolor{black}{$67.63$} & $66.32$ & $66.89$ & $\mathbf{68.49}$ \\ 
\midrule
RAVDESS-Song \cite{livingstone2018ryerson} & $46.05$ & \textcolor{black}{$34.20$} & $9.58$ & \textcolor{black}{$48.48$} & $55.23$ & $56.17$ & $\mathbf{60.11}$ \\ 
\midrule
CaFE \cite{gournay2018canadian} & $52.86$ & \textcolor{black}{$19.22$} & $13.59$ & \textcolor{black}{$53.42$} & $69.23$ & $71.62$ & $\mathbf{76.98}$ \\ 
\midrule
EmoDB \cite{burkhardt2005database} & $62.59$ & \textcolor{black}{$24.77$} & $14.98$ & \textcolor{black}{$66.75$} & $75.42$ & $78.63$ & $\mathbf{83.41}$ \\ 
\midrule \midrule
Avg. & $51.62$ & \textcolor{black}{$41.63$} & $42.23$ & \textcolor{black}{$60.78$} & $62.16$ & $62.81$ & $\mathbf{66.02}$ \\ 
\bottomrule
\end{tabular}}
\vspace{-0.1in}
\end{table*}

\subsection{Comparison with Baselines}
\textcolor{black}{Four baseline systems  (Table~\ref{tab:results-cat}) are considered:-} \\
 \textbf{PASE+:} For each downstream dataset, PASE+ features are extracted and a classification network is trained to predict the emotion class of each utterance similar to \system{}. The total number of parameters used during inference is $8$M. \\
 \noindent\textcolor{black}{\textbf{Whisper:} For each downstream dataset, the representations from the $33$ encoder layers of the Whisper-large-v3 model~\cite{radford2023robust} are  linearly combined with learnable weights. A two-layer classification head is trained on top of these representations for the task of emotion recognition. The total number of parameters used during inference is $800$M.}\\
 \noindent \textbf{Whisper+RoBERTa:}   The transcripts are generated using the Whisper-large-v3 model and subsequently processed by a pre-trained RoBERTa model. The internal representations from RoBERTa are linearly combined by learnable weights, followed by training a two-layer classification head. This has a total of $1.6$B parameters in use during inference.\\
 \noindent \textcolor{black}{\textbf{Teacher-fusion:}   The PASE+ and Whisper+RoBERTa representations are concatenated and a two-layer classification head is trained for each downstream dataset. This baseline also has a total of $1.6$B parameters during inference.}\\
 \textbf{Key takeaways:} 1) The performance of \system{} surpasses that of the acoustic supervisory signal by $29.76\%$ (relative) on average across the $8$ datasets. This improvement can be attributed to the larger parameter size of \system{} compared to the PASE+ model. \textcolor{black}{2) CARE is seen to outperform Whisper and Whisper+RoBERTa systems by $41.79\%$ and $41.18\%$ in relative terms. This indicates that, although the Whisper-based baselines are much larger in size, the combination of the acoustic and semantic information in \system{} results in effective emotion recognition. 3) On MELD and CMU-MOSI, \system{} is outperformed by the Whisper+RoBERTa baseline. For these datasets, text-based models are known to significantly outperform speech-only systems~\cite{dutta2023hcam,lian2022smin}. 
 In the Whisper+RoBERTa setup, the RoBERTa model is fine-tuned on transcripts generated by Whisper-large-v3 ($1.6$B sized model). In contrast, \system{} is a smaller model ($160$M), and does not use directly use the ASR transcripts during inference. To further elucidate the fairness in model-size, 
 we replace Whisper-large-v3 with a Whisper-base model for the ASR, followed by the RoBERTa modeling. Then, the performance drops from $49.29\%$ to $46.02\%$ on MELD and from $75.14\%$ to $71.91\%$ on CMU-MOSI. This underscores the importance of accurate transcriptions and large model capacity in settings where the textual information is emotion rich.} \textcolor{black}{4) While the teacher-fusion baseline is competitive for a number of datasets involving English speech, CARE outperforms this baseline on average by $5.24\%$ absolute. This also motivates why CARE was pre-trained using knowledge distillation as it outperforms the fusion baseline with only $10\%$ of the parameters.}
\subsection{Importance of the Two Encoders}
We present the performance of \system{} when we use only one of acoustic and semantic encoders along with the common encoder for the downstream datasets in Table~\ref{tab:results-cat}. For evaluating the combination of the semantic and common encoders, we use the $768$-dimensional representations from the convolutional feature extractor, the common encoder, and the semantic encoder, excluding outputs from the acoustic encoder. Similarly, the semantic encoder representations are disregarded during the evaluation of the acoustic-common encoder combination. Note that, while \system{} has more number of parameters ($160$M) as compared to models like WavLM or emotion2vec, both these combinations have similar number of parameters during inference. While the semantic-common encoder combination has an inference time parameter size of $110$M, the acoustic-common encoder has a total of $94$M parameters during evaluation on each downstream dataset.\\
\noindent{\textbf{Key takeaways:}} 1) The combination of the acoustic and common encoder representations outperforms the best performing SSL model (HuBERT) by $8.08\%$ (relative) on average for the $8$ datasets (Table~\ref{tab:results-cat-oth}). Given the similar parameter count, this performance suggests an advantage of our pre-training approach. 2) For the three out-of-domain datasets, the acoustic-common combination fares better than its semantic counterpart. 
3) Across all datasets, the combination of both encoders in \system{} yields the highest performance, suggesting that while the individual encoder performances are comparable, they capture distinct characteristics of the speech signal.


\subsection{Modifications in the Semantic Encoder}
To evaluate the suitability of our design choices for the semantic encoder, we made three architectural modifications:
\begin{enumerate}
\item \textbf{\system{}-No init.:} Removing the convolutional adapters, the transformer layers in the semantic encoder are initialized randomly (instead of pre-trained RoBERTa weights). 
\item \textbf{\system{}-Trans.:} Removing the convolutional adapters while the RoBERTa transformer layers are  updated.   
\item \textbf{\system{}-FT:} Keeping the convolutional adapters, we update all the parameters (conv. adapters and transformer weights) in the semantic encoder.
\end{enumerate}
The results for these modifications are shown in Table~\ref{tab:results-cat-ada}. \\
\noindent{\textbf{Key takeaways:}} 1) Initializing the transformer weights with RoBERTa is essential for \system{}'s performance. Random initialization of the semantic encoder leads to performance drops across all five datasets, suggesting that a randomly initialized semantic encoder struggles to regress to the semantic supervisory signal. 2) Removing convolutional adapters (in \system{}-Trans) negatively impacts performance, highlighting the necessity of our convolution-based adaptation technique for aligning speech representations with RoBERTa's transformer layers. 3) Updating the transformer layers in the semantic encoder decreases performance. Since RoBERTa is pre-trained on text, fine-tuning with speech data degrades its effectiveness.
\begin{table}[t!]
\centering
\caption{Results on the downstream datasets (weighted F1-score) with modifications in the semantic encoder.}\label{tab:results-cat-ada}
    \resizebox{0.6\columnwidth}{!}{%
\begin{tabular}{@{}l|ll@{}}
\toprule
Dataset & \multicolumn{1}{l|}{Method} & WF1 \\ \midrule
\multicolumn{1}{l|}{\multirow{3}{*}{IEMOCAP(4-class)}} & \multicolumn{1}{l|}{\system{}-No init.} & $65.71$ \\
\multicolumn{1}{l|}{} & \multicolumn{1}{l|}{\system{}-Trans.} & $65.89$ \\
\multicolumn{1}{l|}{} & \multicolumn{1}{l|}{\system{}-FT} & $68.16$ \\
\multicolumn{1}{l|}{} & \multicolumn{1}{l|}{\system{}} & $\mathbf{69.39}$ \\ \midrule
\multicolumn{1}{l|}{\multirow{3}{*}{IEMOCAP(6-class)}} & \multicolumn{1}{l|}{\system{}-No init.} & $50.96$ \\
\multicolumn{1}{l|}{} & \multicolumn{1}{l|}{\system{}-Trans.} & $51.19$ \\
\multicolumn{1}{l|}{} & \multicolumn{1}{l|}{\system{}-FT} & $52.37$ \\
\multicolumn{1}{l|}{} & \multicolumn{1}{l|}{\system{}} & $\mathbf{55.02}$ \\ \midrule
\multicolumn{1}{l|}{\multirow{3}{*}{MELD}} & \multicolumn{1}{l|}{\system{}-No init.} & $45.16$ \\
\multicolumn{1}{l|}{} & \multicolumn{1}{l|}{\system{}-Trans.} & $46.57$ \\
\multicolumn{1}{l|}{} & \multicolumn{1}{l|}{\system{}-FT} & $47.93$ \\
\multicolumn{1}{l|}{} & \multicolumn{1}{l|}{\system{}} & $\mathbf{48.05}$ \\ \midrule
\multicolumn{1}{l|}{\multirow{3}{*}{CMU-MOSI}} & \multicolumn{1}{l|}{\system{}-No init.} & $62.16$ \\
\multicolumn{1}{l|}{} & \multicolumn{1}{l|}{\system{}-Trans.} & $65.97$ \\
\multicolumn{1}{l|}{} & \multicolumn{1}{l|}{\system{}-FT} & $64.27$ \\
\multicolumn{1}{l|}{} & \multicolumn{1}{l|}{\system{}} & $\mathbf{66.74}$ \\ \midrule
\multicolumn{1}{l|}{\multirow{3}{*}{DAIC-WOZ}} & \multicolumn{1}{l|}{\system{}-No init.} & $64.57$ \\
\multicolumn{1}{l|}{} & \multicolumn{1}{l|}{\system{}-Trans.} & $65.93$ \\
\multicolumn{1}{l|}{} & \multicolumn{1}{l|}{\system{}-FT} & $67.43$ \\
\multicolumn{1}{l|}{} & \multicolumn{1}{l|}{\system{}} & $\mathbf{68.49}$ \\ \bottomrule
\end{tabular}}
\vspace{-0.1in}
\end{table}
\begin{table}[t!]
\centering
\caption{Results on the downstream datasets (weighted F1-score) with different initializations of the acoustic  encoders.}\label{tab:results-ssl}
    \resizebox{\columnwidth}{!}{%

\begin{tabular}{@{}l|c|c|c|c}
\toprule
Datasets & Random init. & HuBERT init. & Data2vec init. &  WavLM init. \\ \midrule
IEMOCAP-4 & $66.72$ &$67.65$ &$66.76$ & $\mathbf{69.39}$  \\ \midrule
IEMOCAP-6 & $51.47$ &$51.40$ &$52.98$  & $\mathbf{55.02}$ \\ \midrule
MELD & $46.41$ &$46.97$ &$47.19$ & $\mathbf{48.05}$  \\ \midrule
CMU-MOSI & $65.07$ &$66.26$ &$\mathbf{68.14}$  & $66.74$ \\ \midrule
DAIC-WOZ & $67.56$ &$65.19$ &$66.59$ & $\mathbf{68.49}$  \\ \midrule
RAVDESS-Song & $57.81$ &$\mathbf{60.30}$ & $55.09$ & $60.11$ \\ \midrule
CaFE & $73.83$ &$72.23$ & $62.46$ & $\mathbf{76.98}$  \\ \midrule
EmoDB & $78.61$ &$\mathbf{86.51}$& $77.19$ & $83.41$ \\ \midrule \midrule
Avg. &$63.44$ &$64.56$ & $62.05$ & $66.02$ \\ \bottomrule

\end{tabular}}
\vspace{-0.1in}
\end{table}

\subsection{Initialization of Acoustic and Common Encoders}
As indicated in Section~\ref{sec:method}, the common and the acoustic encoders of \system{} are initialized with the WavLM-base model weights. We present the results of our method when this initialization is modified to i) random, ii) HuBERT-base~\cite{hsu2021hubert} or iii) data2vec-base~\cite{baevski2022data2vec} (Table~\ref{tab:results-ssl}). \\
\noindent{\textbf{Key takeaways:}} 1) The model's performance decreases with data2vec initialization, likely due to data2vec’s lower baseline performance compared to HuBERT and WavLM (see Table~\ref{tab:results-cat-oth}). An exception is the CMU-MOSI dataset, where this initialization improves over the WavLM initialized model by $4.21\%$ (relative). 2) The HuBERT-initialized model performs best on the RAVDESS-Song and Emo-DB datasets. Notably, HuBERT outperforms WavLM for these two out-of-domain datasets  (Table~\ref{tab:results-cat-oth}). 3) Initialization impacts the acoustic and common encoders less than the semantic encoder, as the latter requires alignment with text representations.

\subsection{Choice of Acoustic Targets}

\begin{figure}
    \centering
    \includegraphics[width=0.6\textwidth,trim={2cm 7cm 5cm 3.5cm},clip]{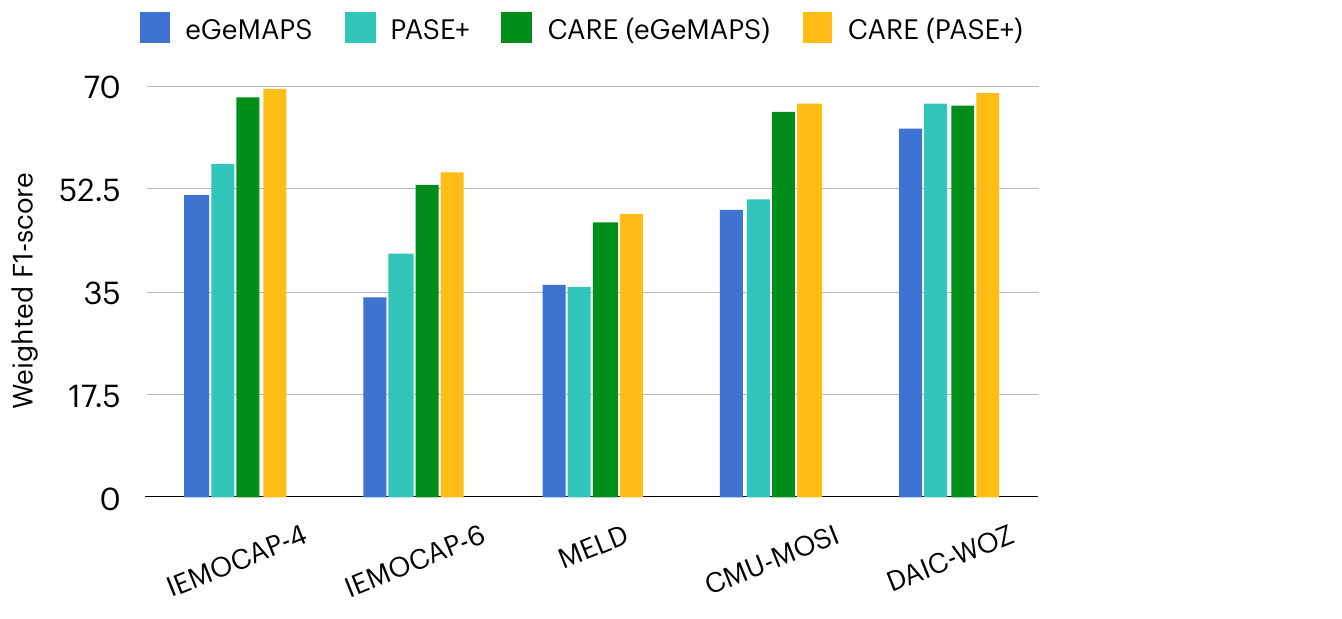}
    \vspace{-0.25in}
    \caption{Performance of \system{} when different acoustic targets are used.  The model with eGeMAPS as features is trained similarly to that of the PASE+ baseline. All numbers are shown as the average of $5$ random initializations.}
    \label{fig:acoustic}
\end{figure}

\begin{figure}
    \centering
    \includegraphics[width=0.6\textwidth,trim={2cm 7cm 5cm 3.5cm},clip]{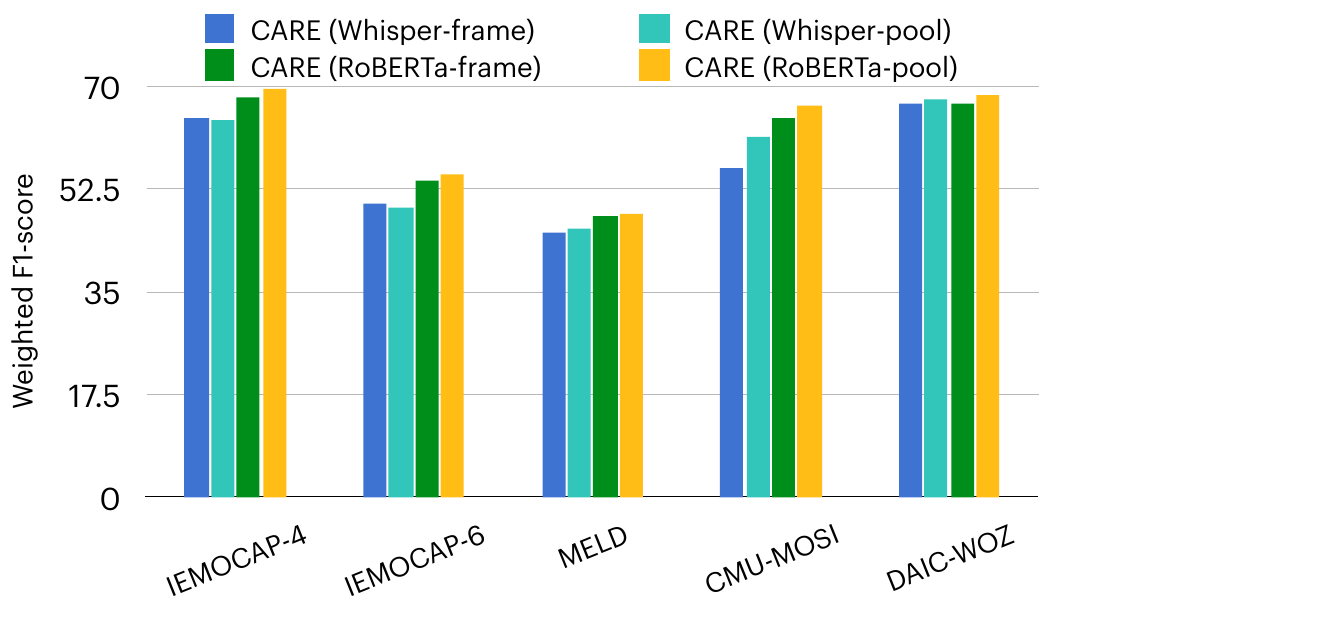}
    \vspace{-0.25in}
    \caption{\textcolor{black}{Performance of \system{} when different semantic targets are used. All numbers are shown as the average of $5$ random initializations.}}
    \label{fig:semantic}
\end{figure}

We run an experiment where the acoustic encoder is trained with targets based on eGeMAPS~\cite{eyben2015geneva} features extracted from the openSMILE toolkit~\cite{eyben2010opensmile}. The  PASE+ targets of the acoustic encoder of the \system{} model is replaced by the eGeMAPS features. The performance of this model, called \system{} (eGeMAPS), is shown in Fig.~\ref{fig:acoustic}. \\
\noindent{\textbf{Key takeaway:}} The baseline model using eGeMAPS input features performs worse than the baseline with PASE+ features, as expected, since eGeMAPS are handcrafted. Consequently, the average performance of CARE with eGeMAPS is also lower than that of CARE with PASE+ targets.
\textcolor{black}{\subsection{Choice of Semantic Targets}}\label{sec:sem_targets}
\textcolor{black}{
We conduct an experiment where the Whisper encoder representations serve as supervisory signals for the semantic encoder. We explore two variants of this: 1) We pool the Whisper representations to serve as semantic targets while pre-training. This model is called \system{} (Whisper-pool). 2) We pre-train a model with the frame-level representations of Whisper as the targets. We call this model \system{} (Whisper-frame). 3) We also use the frame level alignments between speech and the RoBERTa tokens and use the frame-level RoBERTa representations as the semantic targets. We call this model \system{} (RoBERTa-frame). The comparative performances of the different variants along with the proposed model-\system{} (RoBERTa-pool) are shown in Fig.~\ref{fig:semantic}.}\\
\noindent \textcolor{black}{
\noindent{\textbf{Key takeaways:}} 1) The performance of \system{} (RoBERTa-pool) is seen to be superior to both variants trained with Whisper encoded representations.
2) The performance of the systems when the semantic encoder is trained with the pooled   targets is observed to be better than those trained with frame-level representations. }

\begin{figure}[ht!]
    \centering    \includegraphics[width=0.54\textwidth,trim={1.9cm 7.0cm 1cm 3.7cm},clip]{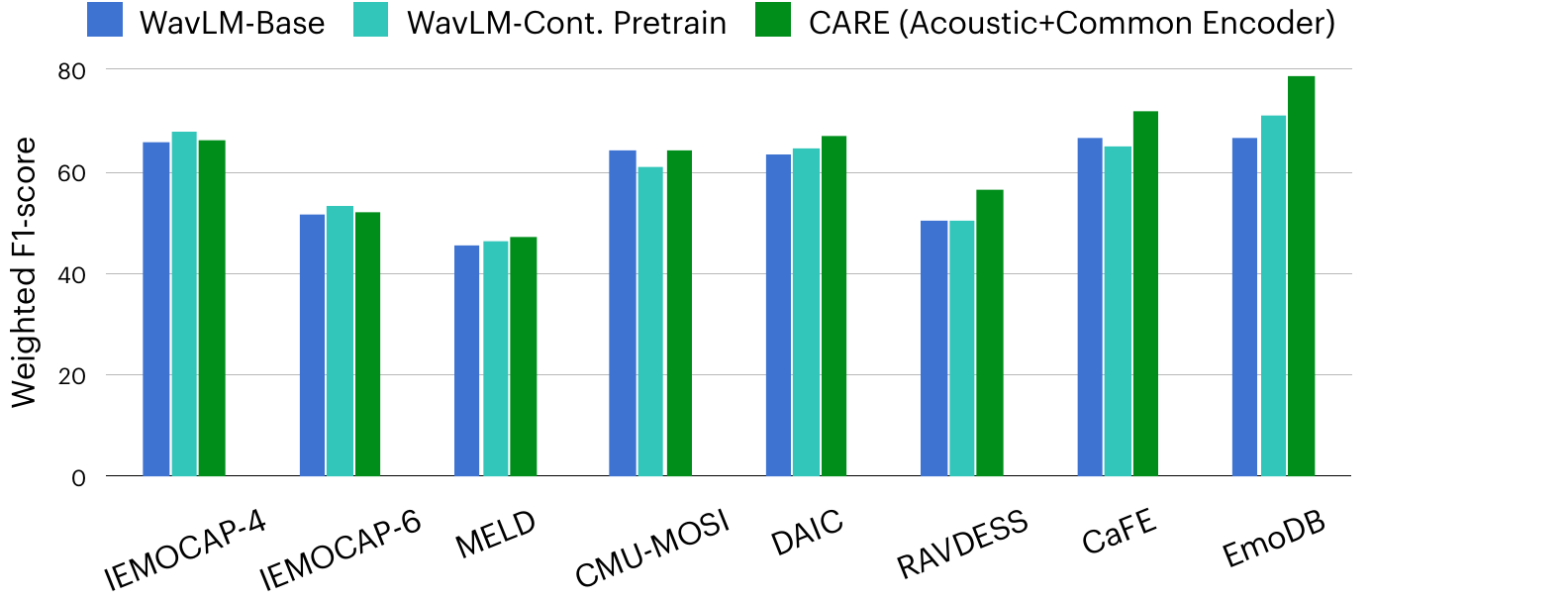}
    \vspace{-0.1in}
    \caption{Comparison of the performance when WavLM is continually pre-trained on MSP-PODCAST. The performance of the combination of acoustic and common encoders of \system{} is shown for reference. Here, RAVDESS refers to the RAVDESS-Song dataset.}
    \label{fig:init}
    \vspace{-0.1in}
\end{figure}

\subsection{Continued Pre-training of WavLM}\label{sec:cont}
Since all self-supervised learning (SSL) models are trained on neutral data, their ability to accurately discern emotions from speech signals is typically limited. The emotion recognition performance of these SSL models when pre-trained on emotion datasets thus becomes crucial. To explore the impact of pre-training setup in the proposed \system{}, we continued the pre-training of the publicly available WavLM-base model,  using the MSP-PODCAST dataset. This was done following the WavLM pre-training procedure, with masked language modeling loss, for an additional $200,000$ steps (similar to the \system{}). The results of this experiment are shown in Fig.~\ref{fig:init}, wherein the performance of the continually pre-trained WavLM model is denoted by {WavLM-Cont. Pretrain}. The performance of the combination of the common and acoustic encoders is also shown for comparison.

\noindent{\textbf{Key takeaway:}} Continued pre-training improves WavLM-base performance on certain downstream tasks, like IEMOCAP. However, except for IEMOCAP, WavLM performs worse than CARE's acoustic-common encoder combination. As all the models in Fig.~\ref{fig:init} have the same size ($94$M) during inference, this experiment highlights the benefits of our proposed distillation-based pre-training.
\textcolor{black}{\subsection{Multimodal Emotion Recognition}}
\textcolor{black}{To assess \system{}'s utility in the multimodal speech-text setting, we design a model using speech (WavLM-base or \system), and text (RoBERTa) fusion. After combining the layer representations in SUPERB style, we concatenate the representations and train a classification head. We   experiment on IEMOCAP-4 and IEMOCAP-6 and observe that the weighted F1-score for CARE+RoBERTa improves from $73.02\%$ to $75.19\%$ for IEMOCAP-4 and from $60.41\%$ to $62.21\%$ for IEMOCAP-6 over WavLM-base+RoBERTa system. 
In addition to the uni-modal improvements reported in Table~\ref{tab:results-cat}, these results highlight that multi-modal speech-text emotion recognition systems can also  benefit from the enhanced representations provided by \system.}
\begin{figure}
    \centering
    \includegraphics[width=\columnwidth,trim={2cm 8cm 10cm 3.5cm},clip]{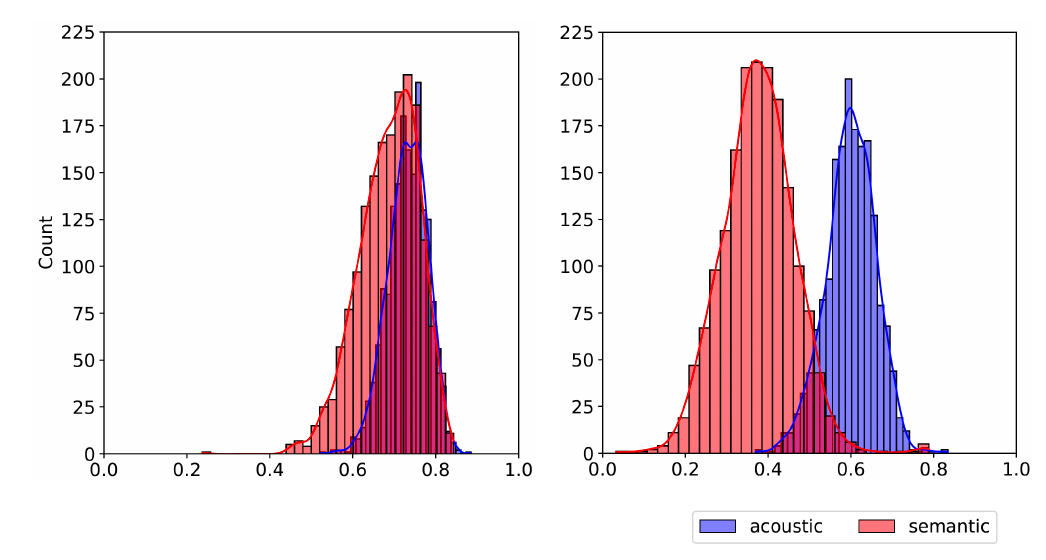}
    \vspace{-20pt}
    \caption{Distribution of cosine similarities for layer $7$ representations for the acoustic and the semantic encoders. The left plot is when two speech signals belonging to \textbf{different speakers, same emotion and same content} are processed by \system{}. The plot on the right refers to the setting with \textbf{different speakers, same emotion and different content}.  
    }
    \label{fig:vis}
\end{figure}

\subsection{Visualization of Layers of \system{}}
In order to interpret the  pre-trained model representations learnt by the acoustic and semantic encoder of \system{}, we probe the  representations from each encoder. We use the English part of the Emotional Speech Dataset (ESD)~\cite{zhou2022emotional} for this analysis. We form pairs of utterances, where both the utterances of a pair have the same emotional label in all cases and they are derived from two different speakers.
A total of $1750$ pairs are considered  and the cosine similarities of the pooled representations (for transformer layer $7$) are shown in  Fig.~\ref{fig:vis}. The figure on the left indicates the setting where the speech content in the two utterances is the same whereas the plot on the right indicates different speech content.  
 We note that when the spoken content is different, the acoustic encoder has higher similarity than the semantic encoder, indicating that the acoustic encoder is beneficial when the emotion information cannot be reliably predicted from the textual content of the audio. 



\section{Summary}
\textbf{Key Highlights:} In this paper, a pre-training technique for content and acoustic encoding of emotional speech is provided. The proposed architecture, termed \system{}, learns an enriched representation of acoustic and semantic information. The acoustic encoder uses supervision from low-level descriptors of speech, while the semantic encoder is distilled using text representations of the speech transcripts. We also propose an adaptation strategy for text-based models in speech representation learning using convolutional neural network layers. The \system{} model, with experiments on $8$ downstream tasks, is seen to outperform  models of comparable sizes on most of the datasets. Further, the \system{} is also observed to generalize better than LLM based models with large parameter sizes. The importance of the different components of the  proposed model, along with the different design choices,  are established through ablation studies. 

\textbf{Limitations and future scope:} The MSP-PODCAST dataset is used for pre-training \system{}, which has only $230$ hours of emotional speech data. 
Another limitation of this work, is the relatively lower performance on some in-domain speech datasets   compared to LLM-based models, like the CMU-MOSI.  In future, we plan to extend the \system{} approach to multi-modal speech-text emotion recognition tasks.

\bibliographystyle{IEEEbib}
\bibliography{refs}

\end{document}